\documentstyle[epsfig,here,12pt]{article}
\begin{document}

\begin{center}
{\Large{{\bf {Comments on the possible discovery of \\
the Higgs boson with mass $\sim$ 160 GeV \\
at the Tevatron\\}}}}

\vskip 15mm
{\small { 
{\bf {Engin ARIK$^1$, Saleh SULTANSOY$^{2,3}$ }}\\
\vskip5mm

$^1$Bo\u gazi\c ci University, Department of Physics, 
34342 Bebek, Istanbul, TURKEY \\
$^2$Gazi University, Department of Physics, 
06500 Teknikokullar, Ankara, TURKEY \\
$^3$Academy of Sciences, Institute of Physics, 
H. Cavid Avenue 33, Baku, AZERBAIJAN}}\\
\end{center}

\vskip10mm

\begin{abstract}
The consequences of the possible discovery of the Higgs boson with a mass
around 160 - 180 GeV at the Tevatron are discussed. 
\end{abstract}

\vskip10mm

Most probably, the search for the Higgs boson will be finalized within few
years. Its discovery will close the large hole in the foundation of the
Standard Model (SM) of elementary particles. At the end of 2000, just before
the LEP collider was shut down, the four experiments (ALEPH, DELPHI, L3, OPAL)
published their search results for the Higgs boson ($H$) signal in the process
$e^+ e^- \rightarrow Z\,H$ at centre-of-mass energies up to 209 GeV. The results
         reported a lower limit of 114.4 GeV on the SM Higgs boson mass with
         95 $\%$ confidence level [1]. Currently, a fit to electroweak
         measurements gives an upper limit of 199 GeV to the preferred mass
         value [2], assuming 3 SM families.

Before the LHC data taking starts in 2007, the upgraded Fermilab Tevatron data 
may signal a discovery. Currently, the collected data amount to $\sim$ 1 fb$^{-1}$ per
experiment which will double in a year. Recently, the two experiments CDF and
D0 at the Tevatron have been searching for a signal in the channel
\begin{eqnarray}
p \bar{p} \rightarrow H\,X \qquad  {\rm{with}} \qquad H \rightarrow WW^{(*)}
\rightarrow l \, \nu\, l' \nu \qquad (l, l' = e, \mu) \,\,.
\end{eqnarray}
The analyzed sample in the above channel corresponds to 
$\sim$ 320 pb$^{-1}$ for D0 and
$\sim$  180 pb$^{-1}$ for CDF [2].  In the Table below, we present the D0 results [3] together
with the addition of the last column indicating the number of signal events
expected with 4 SM families. 

In the recent speculative ('provocative') comments [4] on D0 data [3], it
was pointed out that the data can be interpreted as a clue relating to the
Higgs boson with a mass of $\sim$   160 GeV. Even though there were justified
objections [5] to the way the data was interpreted, the aim of this exercise
was to initiate more activity on the subject. In this study, we clarify our
reasons in attracting attention to this topic and give some suggestions.

As seen in the Table, in the case of 3 SM families, this channel is not
promising for the observation of a Higgs signal at the Tevatron. However, the
Higgs boson can be  discovered before the LHC starts, if the 4$^{th}$ SM family
exists in nature. 

\begin{table}[H]
\begin{center}
{\small Table : 
 Number of signal and background events expected and number of events
observed after all selections are applied. Only statistical uncertainties are
given. We have used an average enhancement factor of 8.5 for the last column.}\\[4mm]
\begin{tabular}{|c|c|c|c|c|}
\hline
m$_H$ (GeV) & $H \rightarrow  WW^{(*)}$ & Background sum & Data & $H \rightarrow  WW^{(*)}$\\  
&  3 families & & & 4 families \\
\hline  \hline
100 & 0.007 $\pm$  0.001 & 31.0 $\pm$ 2.8 & 25 & 0.06 \\
 \hline
120 & 0.11  $\pm$ 0.01 & 30.2  $\pm$ 2.5 & 23 & 0.09 \\
 \hline
140 & 0.33 $\pm$  0.01 & 21.4   $\pm$ 1.6 & 21 & 2.80\\
 \hline
160& 0.54 $\pm$ 0.02 & 17.7  $\pm$ 1.0 & 20 & 4.60\\
 \hline
180 & 0.36 $\pm$  0.01 & 19.1   $\pm$ 1.0  & 20 & 3.06\\
 \hline
200 & 0.17 $\pm$ 0.01 & 20.0  $\pm$ 1.1 & 16 & 1.45\\
 \hline
\end{tabular}
\end{center}
\end{table}

The comparison of the data and the estimated background show a positive excess
region around 160 GeV and deficit at 100, 120 and 200 GeV, which was the
motivation for the comments in [4] where the statistical error on the data was
not taken into account. The above mentioned excess and deficit are not
statistically significant at this level of data. 

The analysis of the present data sample (1 fb$^{-1}$ per experiment) will improve
the statistical significance of the above claim. Let us consider in detail the
case of 160 GeV Higgs at 2 fb$^{-1}$ integrated luminosity. The expected number of
signal (4 SM families) and background events will be $S \sim$ 29 and $B \sim$ 111
respectively, which gives a significance of $S/\sqrt{B}$ = 2.8.

If the two experiments complete the analysis of the already collected data and
join their results they will be able to clarify this issue, otherwise D0 will
have an answer in the next year. In 2006 combined integrated luminosity will be
4 fb$^{-1}$ and 160 GeV mass Higgs could be observed at the Tevatron with an
evidence of $\sim 4\sigma$ . Similar analysis can be extended to the other
                 mass values. For example with 4 fb$^{-1}$ 180 GeV mass Higgs
                 could be observed with an evidence of $\sim 3\sigma$  .

The discovery of the Higgs boson with $m_H$ = 160 - 180 GeV at
                                the Tevatron will change drastically the
                                streamline of high energy
First of all, this will be an indirect manifestation of the 4$^{th}$ SM family. In
this case, the LHC will observe the 4$^{th}$ family quarks directly before
2010. The preferable mass of the 4$^{th}$ family fermions is approximately 600 GeV
which will be a strong argument for the priority of the CLIC as a linear e$^+$ 
e$^-$ collider. The CLIC with $\sqrt{s}$ = 1 - 3 TeV will discover the
   4$^{th}$ family leptons. 

Secondly, this will discourage the SUSY community because the upper limit on
the mass of the Higgs boson in SUSY is $\sim$ 130 GeV. 
The beauty of SUSY should be realized at the most fundamental, pre-preonic,
level as we claimed in [6]. Indeed, the number of observable free parameters
in MSSM is too large [7] to be realistic. 

The LHC will touch the preonic level with large probability. Models with extra
dimensions should also be realized at the most fundamental level therefore the
chance that their predictions will be tested at the LHC is very small.

In conclusion, the existence of the 4$^{th}$ SM family could be confirmed
indirectly in the near future, if $m_H$ is around 160 GeV.

Our suggestions for the Tevatron: 
\begin{itemize}
\item{The analysis of the H '¢ª WW(*) need to be the top priority for both D0 and
CDF experiments.}

\item{D0 and CDF data need to be combined, which will correspond to approximately
  doubling of the integrated luminosity.}
    
\item{A dedicated workshop on this issue before the end of the year will be very desirable.}  
\end{itemize}

\begin{center}
{\bf Acknowledgement }
\end{center}

This work is supported partly by the Turkish State Planning Organization under
the grant number DPT2002K120250 and Turkish Atomic Energy Authority.

\end{document}